\documentclass{article}
\usepackage{dcase2019,amsmath,graphicx,url,times}
\usepackage{color}

\title{Long-distance detection of bioacoustic events \\ with per-channel energy normalization}

\name{
	\!\!\!\!\!\!\!\!\!\!\!\!
	V. Lostanlen$^{1,2}$,
	K. Palmer$^{3}$,
	E. Knight$^{4}$,
	C. Clark$^{1}$,
	H. Klinck$^{1}$,
	A. Farnsworth$^{1}$,
	T. Wong$^{2}$,
	J. Cramer$^{2}$,
	J. Bello$^{2}$}

\address{
$^1$ Cornell Lab of Ornithology, Ithaca, NY, USA\\
$^2$ New York University, Center for Urban Science and Progress, New York, NY, USA\\
$^3$ San Diego State University, San Diego, CA, USA\\
$^4$ University of Alberta, Edmonton, AB, USA}

\usepackage{xspace}

\newcommand*{\ie}{i.e.\@\xspace}

\usepackage{mathptmx}
\usepackage{amsmath}
\usepackage{amssymb}
\usepackage{amsthm}
\newtheorem{thm}{Theorem}[section]

\newtheorem{prop}[thm]{Proposition}
\newtheorem*{prop*}{Proposition}
\theoremstyle{remark}

\usepackage{siunitx}
\usepackage{enumitem}
\usepackage{adjustbox}
\usepackage{stackengine}
\usepackage{svg}
\usepackage{url}
\usepackage{microtype}

\usepackage{lipsum}
\newcommand\blfootnote[1]{%
  \begingroup
  \renewcommand\thefootnote{}\footnote{#1}%
  \addtocounter{footnote}{-1}%
  \endgroup
}

\begin{document}

\ninept
\maketitle

\sloppy

\begin{abstract}
This paper proposes to perform unsupervised detection of bioacoustic events by pooling the magnitudes of spectrogram frames after per-channel energy normalization (PCEN).
Although PCEN was originally developed for speech recognition, it also has beneficial effects in enhancing animal vocalizations, despite the presence of atmospheric absorption and intermittent noise.
We prove that PCEN generalizes logarithm-based spectral flux, yet with a tunable time scale for background noise estimation.
In comparison with pointwise logarithm, PCEN reduces false alarm rate by 50x in the near field and 5x in the far field, both on avian and marine bioacoustic datasets.
Such improvements come at moderate computational cost and require no human intervention, thus heralding a promising future for PCEN in bioacoustics.
\end{abstract}

\section{Introduction}
\label{sec:intro}

The deployment of autonomous recording units offers a minimally invasive sampling of acoustic habitats \cite{shonfield2017ace}, with numerous applications in ecology and conservation biology \cite{efford2009ecology}.
In this context, there is an extensive literature on 
tailoring spectrogram parameters to a specific task of detection or classification: the effects of window size, frequency scale, and discretization are now well understood \cite{ulloa2016screening, knight2019bioacoustics}.
However, the important topic of loudness mapping, \ie{} representing contrast in the time--frequency domain, has received less attention.

This article investigates the impact of distance between sensor and source on the time--frequency representation of acoustic events.
In particular, we point out that measuring local contrast by a difference in pointwise logarithms, as is routinely done in machine learning for bioacoustics, suffers from numerical instabilities in the presence of atmospheric attenuation and intermittent noise.
To address this problem, we propose to employ an adaptive gain control technique known as per-channel energy normalization (PCEN) \cite{wang2017icassp}.

We deliberately err on the side of design simplicity: rather than training a sophisticated classifier, we apply a constant threshold on the time series of max-pooled PCEN magnitudes.
In doing so, our goal is not to achieve the lowest possible false alarm rate, but to argue in favor of replacing the logarithmic mapping of loudness by PCEN in all systems for long-distance sound event detections, including more powerful yet opaque ones such as deep neural networks \cite{zinemanas2019fruct,cartwright2019waspaa}.

Section \ref{sec:pcen} discusses the theoretical benefits of such a replacement: it proves that PCEN extends temporal context beyond a single temporal frame, thus improving effective detection radius.
Sections \ref{sec:coni} and \ref{sec:whale} present applications to avian and underwater bioacoustics respectively, thereby revealing complementary issues: while bird call detection focuses on mitigating atmospheric absorption at high audible frequencies (1--\SI{10}{\kilo\hertz}), whale call detection focuses on mitigating the interference of amplitude-modulated noise from near-field passing ships at low audible frequencies (50--\SI{500}{\hertz}).

\blfootnote{
This work is partially supported by NSF awards 1633206 and 1633259, the Leon Levy Foundation, and the Pinkerton Foundation.

The source code to reproduce experiments and figures is available at: \\
\url{https://www.github.com/BirdVox/lostanlen2019dcase}}

\begin{figure}
\centering
\includegraphics[width=0.9\linewidth]{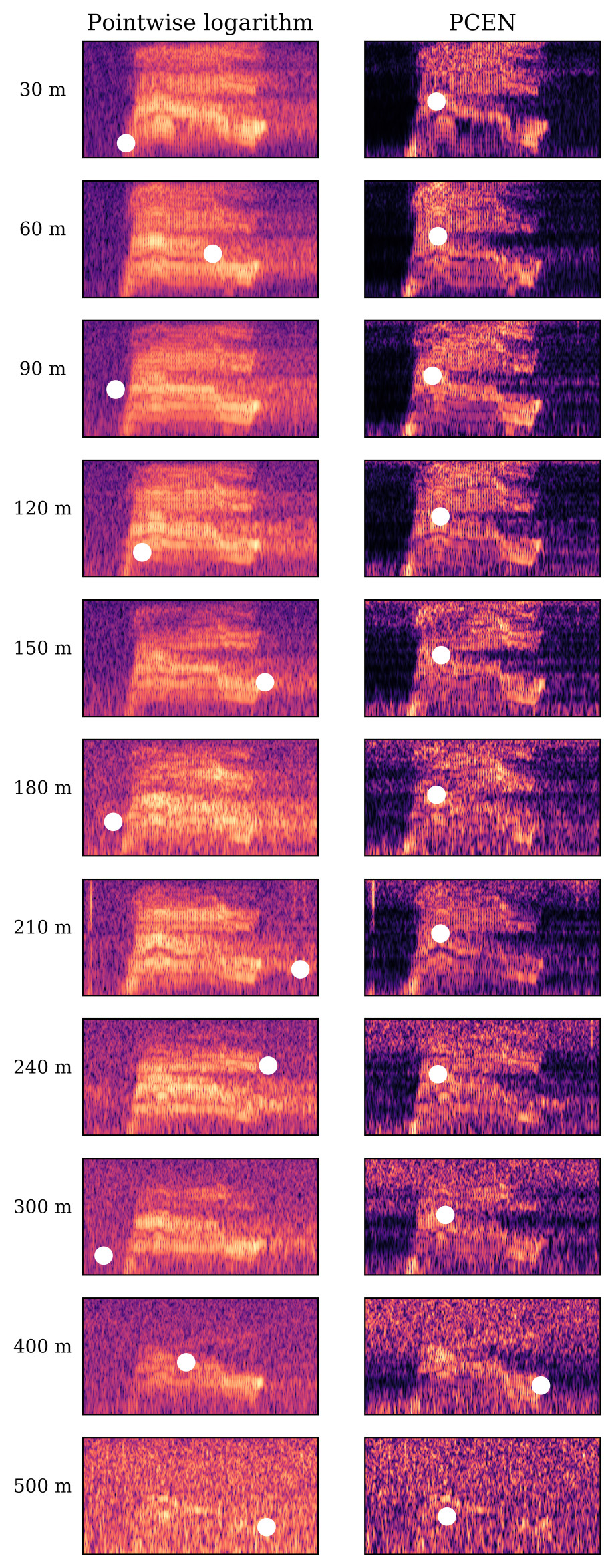}
\caption{Effect of pointwise logarithm (left) and per-channel energy normalization (PCEN, right) on the same Common Nighthawk vocalization, as recorded from various distances.
White dots depict the time--frequency locations of maximal spectral flux (left) or maximal PCEN magnitude (right).
The spectrogram covers a duration of \SI{700}{\milli\second} and a frequency range between 2 and \SI{10}{\kilo\hertz}.
}
\label{fig:coni-pcengrams}
\end{figure}

\section{spectrotemporal measures of novelty}
\label{sec:pcen}

\subsection{Averaged spectral flux}

Let $\mathbf{E}(\mathbf{x})[t,f]$ be the magnitude spectrogram of some discrete-time waveform $\mathbf{x}[t]$.
In full generality, the ordinal variable $f$ may either represent frequency on a linear scale, a mel scale, or a logarithmic scale.
Given $\mathbf{E}$, an implementation of spectral flux composes three operators: loudness mapping, contrast estimation, and feature aggregation.
In its most widespread variant, named \emph{averaged spectral flux}, these three operators respectively correspond to pointwise logarithm, rectified differentiation, and frequential averaging:
\begin{equation}
    \mathrm{SF}_{\mathrm{avg}}(\mathbf{x})[t] =
    \sum_{f}
    \dfrac{
    \max\big(
    \log \mathbf{E}(\mathbf{x})[t,f] - \log \mathbf{E}(\mathbf{x})[t-1, f]
    , 0
    \big)}{N_{\mathrm{fr}}}
    \label{eq:def-sf}
\end{equation}
where $N_{\mathrm{fr}}$ is the number of frequency bands $f$ in $\mathbf{E}$.
The motivation underlying this design choice finds its roots in psychoacoustics, and notably the Weber-Fechner law, which states that the relationship between stimulus and sensation is logarithmic \cite{klapuri1999icassp}.
We may also remark that Equation \ref{eq:def-sf} is invariant to gain.
Indeed, multiplying the waveform $\mathbf{x}[t]$ by some constant $K\neq0$ incurs a multiplication by $K$ in each frequency band of $\mathbf{E}(\mathbf{x})$, and thus an additive bias of $\log K$ in $\log \mathbf{E}(x)$, which eventually cancels after first-order differentiation.
In the case of a single point source at some distance $d$, the relative change in acoustic pressure $K$ caused by a spherical wave propagation is proportional to $\frac{1}{d}$.
Therefore, in a lossless medium without reflections, logarithm-based spectral flux is invariant to geometric spreading insofar as acoustic sources do not overlap in the time--frequency domain.

\subsection{Max-pooled spectral flux}

The situation is different in an absorbing medium.
Indeed, heat conduction and shear viscosity, in conjunction with molecular relaxation processes, attenuate sine waves in quadratic proportion to their frequency \cite{sutherland1998handbook}.
Under standard atmospheric conditions, this attenuation is below \SI{5}{\deci\bel\per\kilo\meter} at \SI{1}{\kilo\hertz}, yet of the order of \SI{100}{\deci\bel\per\kilo\meter} at \SI{10}{\kilo\hertz}.
As a result, bird calls spanning multiple octaves lose in bandwidth as they travel through air.
A simple workaround is to replace the frequential averaging in Equation \ref{eq:def-sf} by a max-pooling operator.
This replacement yields the \emph{max-pooled spectral flux}
\begin{equation}
    \mathrm{SF}_{\max}(\mathbf{x})[t] =
    \max_{f}
    \big(
    \log \mathbf{E}(\mathbf{x})[t,f] - \log \mathbf{E}(\mathbf{x})[t-1, f]
    \big),
\end{equation}
which performs differentiation on a single frequency band, and is thus invariant to the low-pass filtering effect induced by absorption.
However, as illustrated in Figure \ref{fig:coni-pcengrams}, the definition above suffers from numerical instabilities.
Indeed, $\mathrm{SF}_{\max}(\mathbf{x})$ discards all but two scalar values, corresponding to neighboring time--frequency bins in the spectrogram  $\mathbf{E}(\mathbf{x})$.

\subsection{Max-pooled per-channel energy normalization}

In order to associate invariance and stability, this article proposes to increase the time scale of contrast estimation beyond a single spectrogram frame.
To this end, we replace both the logarithmic mapping of loudness and the first-order differentiation by a procedure of per-channel energy normalization (PCEN).
PCEN was recently introduced as a trainable acoustic frontend for far-field automatic speech recognition \cite{wang2017icassp}.
In full generality, PCEN results from an equation of the form
\begin{equation}
    \mathrm{PCEN}(\mathbf{x})[t,f] =
    \dfrac{1}{r}
    \left(
    \dfrac{\mathbf{E}(\mathbf{x})[t,f]}{\big(\varepsilon + \mathbf{M}(\mathbf{x})[t,f] \big)^\alpha} + \delta\right)^{r}
    - \dfrac{\delta^r}{r},
    \label{eq:def-pcen}
\end{equation}
where the gain control matrix $\mathbf{M}(\mathbf{x})$ proceeds from $\mathbf{E}(\mathbf{x})$ by first-order IIR filtering:
\begin{equation}
    \mathbf{M}(\mathbf{x})[t,f] =
    s \times \mathbf{E}(\mathbf{x})[t-1,f]
    +
    (1-s) \times
    \mathbf{M}(\mathbf{x})[t-1,f].
    \label{eq:def-m-exponential}
\end{equation}
Note that the definition in Equation \ref{eq:def-pcen} differs from the original definition \cite{wang2017icassp} by a factor of $\frac{1}{r}$. This is in order to allow the limit case $r\rightarrow0$ to remain nonzero.
Investigating the role of all parameters in PCEN is beyond the scope of this paper; we refer to the asymptotic analysis of \cite{lostanlen2018spl} in this regard.
Rather, we focus on the smoothing parameter $0<s<1$ as striking a tradeoff between numerical stability ($s\rightarrow0$) and rapid adaptation to nonstationary in background noise ($s\rightarrow1$).
The following proposition, proven in Section \ref{sec:appendix}, asserts that PCEN is essentially a generalization of spectral flux.
\begin{prop}
\label{prop:pcen-limit}
At the limit $(s, \varepsilon, \alpha, r) \rightarrow (1, 0, 1,  0)$ in Equations \ref{eq:def-pcen} and \ref{eq:def-m-exponential}, and for any finite value of $\delta$, $\mathrm{PCEN}(\mathbf{x})(t,f)$ tends towards
\begin{equation}
\log\big(\mathbf{E}(\mathbf{x})[t,f] + \mathbf{E}(\mathbf{x})[t-1,f]\big) - \log\mathbf{E}(\mathbf{x})[t-1,f],
\label{eq:sf-softplus}
\end{equation}
which is a smooth approximation of the summand in Equation \ref{eq:def-sf}.
\end{prop}
For the sake of simplicity, we adopt the PCEN parametrization that is prescribed by Proposition \ref{prop:pcen-limit}: we set $\varepsilon=0$, $\alpha=1$, $\delta=1$, and $r=0$.
Derecursifying the autoregressive dependency in \ref{eq:def-m-exponential} and summarizing across frequencies yields the \emph{max-pooled PCEN} detection function
\begin{align}
    \mathrm{PCEN}_{\max}(\mathbf{x})[t] =
    \log
    \Bigg(
    1 +
    \max_{f} 
    \dfrac{\mathbf{E}(\mathbf{x})[t,f]}{s \sum_{\tau=0}^{+\infty} (1-s)^{\tau} \mathbf{E}(\mathbf{x})[t-\tau-1,f]}\Bigg).
\end{align}

\section{Application to avian bioacoustics}
\label{sec:coni}

\subsection{CONI-Knight dataset of Common Nighthawk calls}

We consider the problem of detecting isolated calls from breeding birds in a moderately cluttered habitat.
To this end, we use the CONI-Knight dataset \cite{knight2018bioacoustics}, which contains $64$ vocalizations from five different adult male Common Nighthawks (\emph{Chordeiles minor}), as recorded by $11$ autonomous recording units in a regenerating pine forest north of Fort McMurray, AB, CA.
The acoustic sensor network forms a linear transect, in which the distance between each microphone and the vocalizing individual varies from \SI{30}{\meter} to \SI{500}{\meter}.
The dataset contains $11\times64 = 704$ positive audio clips in total, each lasting \SI{700}{\milli\second}.
These clips were annotated by an expert, as part of a larger collection of continuous recordings which lasts seven hours in total.
We represent each of these clips by their mel-frequency magnitude spectrograms, consisting of $128$ bands between \SI{2}{\kilo\hertz} and \SI{11.025}{\kilo\hertz}, and computed with a Hann window of duration \SI{12}{\milli\second} ($256$ samples) and hop \SI{1.5}{\milli\second} ($32$ samples).
These parameters are identical as in the state-of-the-art deep learning model for bird species recognition from flight calls \cite{salamon2017fusing}.
We use the librosa implementation of PCEN \cite{mcfee2019librosa} with $s=0.09$, \ie{} an averaging time scale of about $T=\SI{100}{\milli\second}$.

Figure \ref{fig:coni-pcengrams} displays the mel-frequency spectrogram of one call at various distances, after processing them with either pointwise logarithm (left) or PCEN (right).
Atmospheric absorption is particularly noticeable above \SI{200}{\meter}, especially in the highest frequencies.
Furthermore, we observe that max-pooled spectral flux is numerically unstable, because it triggers at different time--frequency bins from one sensor to the next.
In comparison, PCEN is more consistent in reaching maximal magnitude at the onset of the call, and at the same frequency band.

\begin{figure}[t]
\centering
\includegraphics[width=\linewidth]{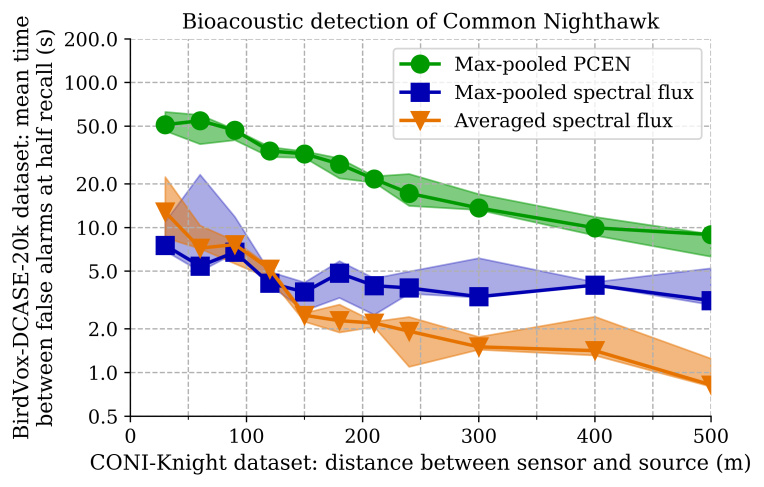}
\caption{Detection of Common Nighthawk calls: evolution of mean time between false alarms at half recall (MTBFA@50) as a function of distance between sensor of source. Shaded areas denote interquartile variations across individual birds. See Section \ref{sec:coni} for details.}
\label{fig:mtbfa50-coni}
\end{figure}

\subsection{Evaluation: mean time between false alarms at half recall}

Our evaluation procedure consists in two stages: distance-specific threshold calibration and estimation of false alarm rate.
In the first stage, we split the dataset of positive clips (\ie{} containing one vocalization) into disjoint subsets of increasing average distance; sort the values of the detection function over this subset in decreasing order; and set the detection threshold at the median value, thus yielding a detection recall of $50\%$.
In the second stage, we run the detector on an external dataset of negative recordings, \ie{} containing no vocalizations from the species of interest; apply the detection thresholds that were prescribed by the first stage; and count the number of false alarms, \ie{} values of the detection function that are above threshold.
Dividing the total duration of the dataset of negative recordings by this number of peaks above threshold yields the \emph{mean time between false alarms at half recall} (MTBFA@50) of the detector, which grows in inverse proportion to false alarm rate.
We repeat this operation over all available subsets to obtain a curve that decreases with distance, and which reflects the ability of the detection curve to generalize from near-field to far-field events.

\begin{figure}[t]
\centering
\includegraphics[width=\linewidth]{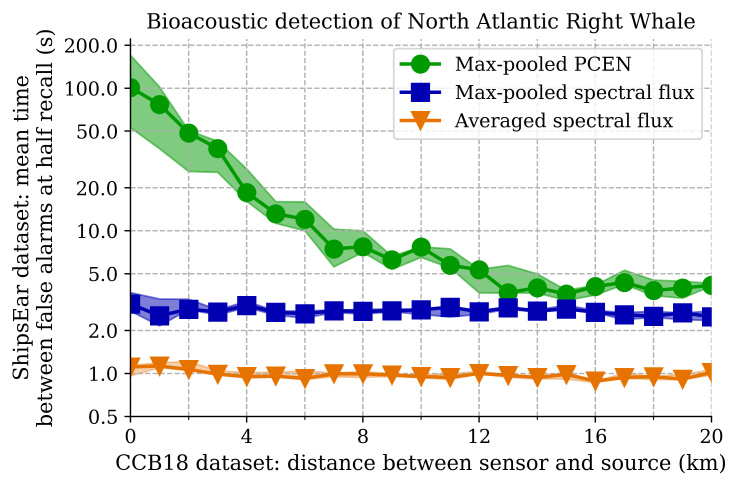}
\caption{Detection of North Atlantic Right Whale calls: evolution of mean time between false alarms at half recall (MTBFA@50) as a function of distance between sensor of source. Shaded areas denote interquartile variations across days. See Section \ref{sec:whale} for details.}
\label{fig:mtbfa50-whale}
\end{figure}

\subsection{Results and discussion}

In the case of the Common Nighthawk, we choose the BirdVox-DCASE-20k dataset \cite{lostanlen2018bvdcase20k} as a source of negative recordings.
A derivative of BirdVox-full-night \cite{lostanlen2018icassp}, this dataset has been divided into $20$k ten-second soundscapes from six autonomous recording units in Ithaca, NY, USA, and annotated by an expert for presence of bird calls.
Among these $20$k soundscapes, $9983$ are guaranteed to contain no bird call, and \emph{a fortiori} no Common Nighthawk call.
These $9983$ recordings amount to 27 hours of audio, \ie{} over 30M spectrogram frames.
For each detection function, we subtract the minimum value over each $10$-second scene to the frame-wise value, in order to account for the nonstationarity in background noise at the scale of multiple hours.

Figure \ref{fig:mtbfa50-coni} summarizes our results.
We find that max-pooled PCEN enjoys a five-fold reduction in false alarm rate with respect to average spectral flux.
In addition, the false alarm rate at \SI{300}{\meter} of max-pooled PCEN is comparable with the false alarm rate of averaged spectral flux at \SI{30}{\meter}.
As a post hoc qualitative analysis, we compute novelty curves for $200$ recordings of outdoor noise from the ESC-50 dataset \cite{piczak2015mlsp}: geophony (rain, wind), biophony (crickets), and anthropophony (helicopter, chainsaw).
For max-pooled spectral flux, we find that the main causes of false alarms are pouring water ($30$\% of total amount), crackling fire ($17$\%), and water drops ($10$\%).

\section{Application to marine bioacoustics}
\label{sec:whale}

\subsection{CCB18 dataset of North Atlantic Right Whale calls}

We consider the problem of detecting isolated calls from whales in a noisy environment.
To this end, we use the CCB18 dataset, which contains vocalizations from about 80 North Atlantic Right Whales (\emph{Eubalaena glacialis}), as recorded by nine underwater sensors during five days in Cape Cod Bay, MA, USA.
The distance between sensor and source is estimated by acoustic beamforming, similarly as in \cite{clark2019jcrm}.
The dataset contains 40k clips in total, each lasting two seconds.
These clips were annotated by an expert, as part of a larger collection of continuous recordings which lasts 1k hours in total.
We represent each of these clips by their short-term Fourier transform (STFT) magnitude spectrograms, consisting of $128$ bands between \SI{8}{\hertz} and \SI{1}{\kilo\hertz}, and computed with a Hann window of duration \SI{128}{\milli\second} and hop of \SI{64}{\milli\second}.
We set $s=0.33$, \ie{} an averaging time scale of about $T=\SI{1}{\second}$.
We choose the ShipsEar dataset as a source of negative recordings \cite{santos2016appliedacoustics}.
This dataset contains 90 ship underwater noise recordings from vessels of various sizes, most of them acquired at a distance of \SI{50}{\meter} or less.
These 90 recordings amount to 189 minutes of audio, \ie{} 177k spectrogram frames.

\subsection{Results and discussion}

Figure \ref{fig:mtbfa50-whale} summarizes our results.
First, we find that averaged spectral flux leads to poor false alarm rates, even in the near field.
We postulate that this is because, in the CCB18 dataset, ship passage events occasionally introduce high received levels of noise.
In other words, distance sets an upper bound, but no lower bound, on signal-to-noise ratio.
Therefore, achieving $50\%$ recall with averaged spectral flux requires to employ a low detection threshold, which in turn triggers numerous false alarms.

Secondly, we find that, across the board, replacing averaged spectral flux by max-pooled spectral flux allows a two-fold reduction in false alarm rate.
We postulate that this improvement is due to the fact that whale calls are locally sinusoidal whereas near-field ship noise is broadband.
Indeed, the max-pooled spectral flux of a chirp is above its averaged spectral flux, with a ratio of the order of $N_{\mathrm{fr}}$; whereas the averaged and max-pooled spectral fluxes of a Dirac impulse are the same.
Therefore, maximum pooling is particularly well suited to the extraction of chirps in noise \cite{bock2013dafx}.

Thirdly, we find that, in the near field, replacing spectral flux by PCEN leads to a $50$-fold reduction in false alarm rate.
We postulate that this is because ship noise has rapid amplitude modulations, at typical periods of 50 to \SI{500}{\milli\second} (\ie{} engine speeds of 120 to 1200 rotations per minute).
If this period approaches twice the hop duration (\ie{} \SI{128}{\milli\second} in our case), short-term magnitudes $\log \mathbf{E}(x)[t-1,f]$ and $\mathbf{E}(x)[t,f]$ may correspond precisely to intake and expansion in the two-stroke cycle of the ship, thus eliciting large values of spectral flux.
Nevertheless, in the case of PCEN, the periodic activation of one every other frame causes $\mathbf{M}(\mathbf{x})[t,f]$ to be of the order of $\frac{1}{2}\mathbf{E}(\mathbf{x})[t,f]$, assuming that the parameter $T$ is large enough to encompass multiple periods.
Therefore, $\mathrm{PCEN}_{\max}(\boldsymbol{x})[t]$ peaks at $\log(\frac{3}{2})$ in the absence of any transient signal.
This peak value is relatively low in comparison with the max-pooled PCEN of a near- or mid-field whale call.

Fourthly, we find that the false alarm rate of max-pooled PCEN increases exponentially with distance, until reaching comparable values as max-pooled spectral flux at a distance of \SI{12}{\kilo\meter}.
This decay is due, in part, to geometric spreading, but also to more complex acoustic phenomena, such as reflections and scattering with the surface as well as the ocean floor \cite{hodges2010book}.
At these large distances, a successful detector should not only denoise, but also dereverberate whale calls.
Max-pooled PCEN does not have any mechanism for dereverberation, and thus falls short of that objective.
Thus, an ad hoc detection function is no longer sufficient, and the resort to advanced machine learning techniques appears as necessary.
We must note, however, that deep convolutional networks in the time--frequency domain rely on the same functional blocks as max-pooled PCEN --- \ie{} rectified extraction of local contrast and max-pooling --- albeit in a more sophisticated, data-driven fashion.
Consequently, we believe that PCEN, whether parametrized by feature learning or by domain-specific knowledge, has a promising future in deep learning for environmental bioacoustics.

\section{Conclusion}

An adequate representation of loudness in the time--frequency domain is paramount to efficient sound event detection.
This is particularly true in bioacoustic monitoring applications, where the source of interest may vocalize at a large distance to the microphone.
Our experiments on the Common Nighthawk and the North Atlantic Right Whale demonstrate that, given a simple maximum pooling procedure across frequencies, per-channel energy normalization (PCEN) outperforms conventional (logarithm-based) spectral flux.
Beyond the direct comparison between ad hoc detection functions at various distances, this study illustrates the appeal in replacing pointwise logarithm by PCEN in time--frequency representations of mid- and far-field audio signals.
In the future, PCEN could be used, for example, as a similarity measure for spectrotemporal template matching; as an input to deep convolutional networks in the time--frequency domain \cite{lostanlen2019plosone}; or as a frequency-dependent acoustic complexity index for visualizing nonstationary effects in ``false color spectrograms'' \cite{towsey2014procedia} of open soundscapes.

\section{Appendix: proof of Proposition 2.1}
\label{sec:appendix}

\begin{proof}
Applying Taylor's theorem to the exponential function yields
\begin{equation}
    \dfrac{\delta^r}{r}
    \Bigg[
    \Big(1 + \frac{\mathbf{E}(\mathbf{x})[t,f])}{\mathbf{M}(\mathbf{x})[t,f]}\Big)^r - 1
    \Bigg]
    \approx
    \delta^r \log\Big(1 + \dfrac{\mathbf{E}(\mathbf{x})[t,f]}{\mathbf{M}(\mathbf{x})[t,f]} \Big)
\end{equation}
with an error term proportional to $r \delta^r \log(1 + \frac{\mathbf{E}(\mathbf{x})[t,f]}{\mathbf{M}(\mathbf{x})[t,f]})^2$, which vanishes at the limit $r \rightarrow 0$ as long as $\mathbf{M}(\mathbf{x})[t,f]$ remains nonzero.
On the left-hand side, we recognize $\mathbf{PCEN}(\mathbf{x})[t,f]$ with $\varepsilon = 0$ and $\alpha = 1$.
On the right-hand side, the finite factor $\delta^r$ tends towards $1$ for $r\rightarrow0$.
The limit $s\rightarrow1$ allows to replace $\mathbf{M}(\mathbf{x})[t,f]$ by $\mathbf{E}(\mathbf{x})[t-1,f]$.
We conclude with
\begin{align}
    \log\Big(1 + \dfrac{\mathbf{E}(\mathbf{x})[t,f]}{\mathbf{E}(\mathbf{x})[t-1,f]} \Big)
    =
    \log\Big( \dfrac{\mathbf{E}(\mathbf{x})[t-1,f] + \mathbf{E}(\mathbf{x})[t,f]}{\mathbf{E}(\mathbf{x})[t-1,f]} \Big) \nonumber \\
    =
    \log\big(\mathbf{E}(\mathbf{x})[t,f] + \mathbf{E}(\mathbf{x})[t-1,f]\big) 
    -
    \log\mathbf{E}(\mathbf{x})[t-1,f].
\end{align}
Interestingly, the distinction between Equation \ref{eq:def-sf} and Equation \ref{eq:sf-softplus} mirrors the distinction between the rectified linear unit (ReLU) $y \mapsto \max(y, 0)$ and the softplus $y \mapsto \log(1 + \exp(y))$ in deep learning.
\end{proof}

\section{Acknowledgment}
We wish to thank D. Santos-Dom\'{i}nguez for sharing his dataset.

\bibliographystyle{IEEEtran}
\bibliography{refs19}

\end{document}